\documentstyle[psfig]{mn}
\title{The rms peculiar velocity of galaxy clusters for different
cluster masses and radii}

\author[Ivan Suhhonenko and Mirt Gramann]
{Ivan Suhhonenko and Mirt Gramann
   \\
   Tartu Observatory,
       T\~oravere 61602, Estonia}
\begin{document}
\maketitle

\let\sec=\section
\let\ssec=\subsection
\let\sssec=\subsubsection

\def\kms{\;{\rm km\,s^{-1}}}
\def\kmsmpc{\;{\rm km\,s^{-1}\,Mpc^{-1}}}
\def\hompc{\,h\,{\rm Mpc}^{-1}}
\def\mpcoh{\,h^{-1}\,{\rm Mpc}}
\def\mpc3h{\,h^{3}\,{\rm Mpc^{-3}}}

\begin{abstract}
We investigate the rms peculiar velocity of galaxy clusters in
the Lambda cold dark matter ($\Lambda$CDM) and tau cold dark
matter ($\tau$CDM) cosmological models using N-body simulations.
Cluster velocities for different cluster masses and radii are
examined. To identify clusters in the simulations we use two
methods: the standard friends-of-friends (FOF) method and the
method, where the clusters are defined as the maxima of the
density field smoothed on the scale $R\sim 1h^{-1}$ Mpc
(DENSMAX). If we use the DENSMAX method, the size of the selected
clusters is similar for all clusters. We find that the rms
velocity of clusters defined with the DENSMAX method is almost
independent of the cluster density and similar to the linear
theory expectations. The rms velocity of FOF clusters decreases
with the cluster mass and radius. In the $\Lambda$CDM model, the
rms peculiar velocity of massive clusters with an intercluster
separation $d_{cl}=50h^{-1}$ Mpc is $\approx$15\% smaller than
the rms velocity of the clusters with a separation
$d_{cl}=10h^{-1}$Mpc.
\end{abstract}


\sec{INTRODUCTION}

One of the interesting unknowns in cosmology is the large-scale
peculiar velocity field in the Universe. The peculiar velocity
field can be studied by using galaxies or clusters of galaxies.
However, there are some advantages in studying the peculiar
velocity field by using galaxy clusters. One of these advantages
comes from the fact that, on scales probed by galaxy clusters,
velocity fluctuations are largely in the quasi-linear regime
and close to the initial state from which large scale
structures developed. In addition, peculiar velocities of
clusters can be determined more accurately than peculiar
velocities of galaxies since the distance to each cluster can be
obtained from a large number of member galaxies, thus
considerably reducing the velocity uncertainties of clusters.
Cluster motions could therefore provide an important tool in
probing the large-scale peculiar velocity field.

Peculiar velocities of clusters of galaxies have been studied
in several papers (e.g. Bahcall, Gramann \& Cen 1994; Lauer \&
Postman 1994; Bahcall \& Oh 1996; Moscardini et al. 1996; Borgani
et al. 1997; Watkins 1997; Dale et al. 1999; Hudson et al. 1999;
Borgani et al. 2000; Colless et al. 2001). Watkins (1997)
developed a likelihood method for estimating the rms peculiar
velocity of clusters from line-of sight velocity measurements.
This method was applied to two observed
samples of cluster peculiar velocities: the SCI
sample (Giovanelli et al. 1997) and a subsample of the Mark III
catalogue (Willick et al. 1997). Watkins (1997) found that the
rms one-dimensional cluster peculiar velocity is
$256^{+106}_{-75}$ km s$^{-1}$, which corresponds to the
three-dimensional rms velocity $459^{+184}_{-130}$ km s$^{-1}$ .
Dale et al. (1999) obtained Tully-Fisher peculiar velocities for
52 Abell clusters distributed over the whole the sky between $\sim
50$ and $\sim 200h^{-1}$ Mpc. They found that the rms
one-dimensional cluster peculiar velocity is $341 \pm 93$ km
s$^{-1}$, which corresponds to the three-dimensional rms velocity
$591 \pm 161$ km s$^{-1}$.

Radial peculiar velocities of clusters can be determined to large
distances by measuring the Sunyaev-Zeldovich  (1980) (SZ)
effect. Rephaeli \& Lahav (1991) made one of the first estimates
of the possibility of measuring the peculiar velocities by using
the SZ effect for a selected sample of galaxy clusters. However,
most convincing measurements for individual clusters have been
done only recently, using the new generation of sensitive
bolometers (Holzapfel et al. 1997; Lamarre et al. 1998). The
accuracy of SZ measurements for determining peculiar velocities
of clusters have been studied in several papers (Haehnelt \&
Tegmark 1996; Aghanim, Gorski \& Puget 2001; Diego et al. 2002;
Holder 2002; Nagai, Kravtsov \& Kosowsky 2002). With $\mu K$
sensitivity on arcminute scales at several frequencies it will be
possible to measure peculiar velocities to an accuracy of
$\sim$130 km s$^{-1}$.

In this paper we study the rms peculiar velocity of clusters,
$v_{rms}$, in different cosmological models assuming that the initial density
fluctuation field is a Gaussian field. To investigate the
nonlinear regime, we use N-body simulations. We examine
cluster peculiar velocities for different cluster masses.
Do cluster velocities depend on their masses? The rms peculiar velocity of
peaks in the initial Gaussian field does not depend on the height of peaks
(Bardeen et al. 1986). However, as an initially Gaussian density field evolves
gravitationally in the nonlinear regime, it becomes
non-Gaussian.

The evolution of peculiar velocities of galaxy clusters in
different N-body models with an initially Gaussian density field
has been examined in several papers (e.g. Bahcall, Gramann, Cen
1994; Croft \& Efstathiou 1994; Suhhonenko \& Gramann 1999,
Colberg et al 2000; Sheth \& Diaferio 2001). Groth \& Efstathiou
(1994) studied the cluster peculiar velocities for two different
cluster richnesses, described
by the mean intercluster separations $d_{cl}=30h^{-1}$ Mpc
and $d_{cl}=55h^{-1}$ Mpc. They found that the peculiar velocity
distributions are almost independent of cluster richness. Sheth
\& Diaferio (2001) studied rms peculiar velocity of clusters
for different masses and found that in N-body simulations the rms
velocity of clusters depends weakly on cluster mass, with a
small trend that for more massive clusters the rms peculiar
velocity decreases. On the other hand, Suhhonenko \& Gramann
(1999) investigated the properties of clusters using N-body
simulations and found that the rms peculiar velocities of
clusters increase with cluster richness.

In this paper we study the dependence of $v_{rms}$ on cluster
masses in more detail. We also examine the rms peculiar
velocities of clusters for different cluster radii. We use the
N-body simulations published by the Virgo Consortium and
described in detail by Jenkins et al. (1998). These simulations
were carried out using a parallel, adaptive
particle-particle/particle-mesh (AP$^3$M) code (Couchman, Thomas
\& Pearce 1995; Pearce \& Couchman 1997). In this paper we
analyze the velocities in the $\Lambda$CDM and $\tau$CDM model
(see Jenkins et al. (1998) and Section 2 for the description of
the cosmological parameters in these models). We also follow the
evolution of particles in a similar $\Lambda$CDM model with the
same cosmological parameters but using a particle-mesh (PM) code.
In this PM code we use a traditional two-point finite-difference
approximation to calculate the forces on the grid. We also
examine the shifted-mesh scheme (see eq. (13) below) and show that this
scheme artificially boosts the velocities of clusters.

To identify clusters in the simulations we use two methods: the
standard friends-of-friends (FOF) method and the method, were the
clusters are defined as the maxima of the density field smoothed
on the scale $R\sim 1h^{-1}$ Mpc (DENSMAX). To determine the
velocity of DENSMAX clusters, we use the same smoothing scale as
for the density field. The velocity of FOF clusters is defined to
be the mean velocity of all the particles in the cluster.

This paper is organized as follows. In Section 2 we describe the
cosmological models, N-body simulations and cluster selection
algorithms used. We also examine the linear theory predictions
for the peculiar velocities of peaks in the Gaussian field. In
Section 3 we examine the rms peculiar velocity of clusters for
different cluster masses and radii, and compare the cluster
velocities with the linear theory predictions. In Section~4 we briefly
discuss the effect of the shifted-mesh scheme on the cluster
velocities. A summary and discussion are presented in Section~5.

\sec{MODELS}

\ssec{Simulations}

We analyze peculiar velocities in N-body simulations
carried out by the Virgo consortium for two cosmological models,
$\Lambda$CDM ($\Omega_0=0.3$, $\Lambda=0.7$) and $\tau$CDM
($\Omega_0=1)$ as described by Jenkins et al. (1998). In these
cold dark matter (CDM) models the power spectrum of the initial
conditions was chosen to be in the form given by Bond \& Efstathiou (1984),
$$
P(k)={Ak \over [1+(aq+(bq)^{3/2}+(cq)^2)^\nu]^{2/\nu}}, \eqno(1)
$$
where $q=k/\Gamma$, $a=6.4h^{-1}$ Mpc, $b=3h^{-1}$ Mpc,
$c=1.7h^{-1}$ Mpc, $\nu=1.13$ and $\Gamma=0.21$. The
normalization constant, $A$, was chosen by fixing the value of
$\sigma_8$ (the linearly extrapolated mass fluctuation in spheres
of radius $8h^{-1}$ Mpc); $\sigma_8=0.9$ and $\sigma_8=0.51$
for the $\Lambda$CDM and $\tau$CDM model, respectively.

We investigated the linear theory predictions for peculiar
velocities of peaks in the $\Lambda$CDM and $\tau$CDM model. The
linear rms velocity fluctuation on a given scale $R$ can be
expressed as
$$ \sigma_v(R)=H_0 f(\Omega_0) \sigma_{-1} (R), \eqno(2)
$$
where the spectral moments $\sigma_j$ are defined for any integer $j$ by
$$
\sigma_j^2={1 \over 2\pi^2} \int P(k) W^2(kR) k^{2j+2} dk,
\eqno(3)
$$
$W(kR)$ is a window function and $f(\Omega_0)$ is the dimensionless
growth rate. The function $f(\Omega_0)=0.51$ and $f(\Omega_0)=1.0$ in the
$\Lambda$CDM and $\tau$CDM model, respectively. (We note that the
approximation $f(\Omega_0)=\Omega_0^{0.6}$ underestimates the
dimensionless growth rate by $\sim 5$\% in the flat
$\Omega_0=0.3$ model).

Bardeen et al. (1986) showed that in the linear approximation the rms peculiar
velocity at peaks of the smoothed density field differs systematically from
$\sigma_v(R)$, and can be expressed as
$$
\sigma_p(R)=\sigma_v(R) \sqrt{1 - \sigma_0^4/\sigma_1^2
\sigma_{-1}^2} . \eqno(4)
$$
In this approximation, the rms velocities of peaks do not
depend on the height of the peaks.

\begin{figure}
\centering
\leavevmode
\psfig{file=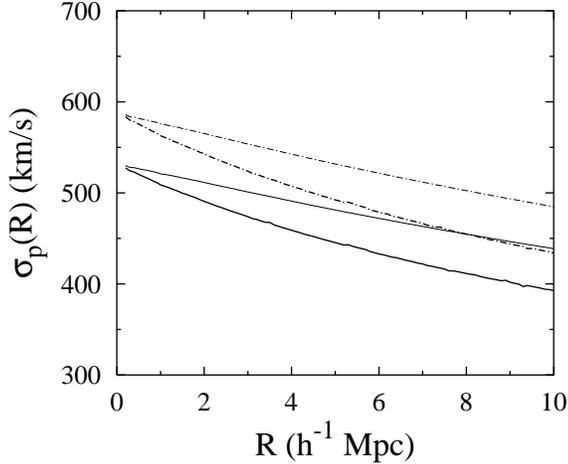,width=8cm}

\caption {The rms peculiar velocity of peaks, $\sigma_p(R)$, in
the $\Lambda$CDM model with $\Omega_0=0.3$ and $\sigma_8 = 0.9$
(heavy solid line) and in the $\tau$CDM model with $\Omega_0=1.0$
and $\sigma_8 = 0.51$ (heavy dot-dashed line). The light curves
show the corresponding rms peculiar velocity $\sigma_v(R)$ for
the same models.}
\end{figure}

Fig. 1 shows the rms peculiar velocity of peaks, $\sigma_p(R)$,
for the $\Lambda$CDM and $\tau$CDM model. We have used the
top-hat window function. For comparison, we show also the rms
peculiar velocity $\sigma_v(R)$ for the same models. For the
radius $R=1h^{-1}$ Mpc, $\sigma_p=509$ km/s and $\sigma_p=562$ km/s
in the $\Lambda$CDM and $\tau$CDM models, respectively.
At this radius, $\sigma_p$ is lower than $\sigma_v$ about
$\sim 2$ per cent for the models studied. On larger scales, the
difference between $\sigma_p$ and $\sigma_v$ increases.

Next we study peculiar velocities in N-body simulations.
The Virgo simulations were created using a parallel adaptive
particle-particle/particle-mesh (AP$^3$M) code as described by
Couchman, Thomas \& Pearce (1995) and Pearce \& Couchman (1997).
It supplements the standard P$^3$M algorithm (Efstathiou et al.
1985) by recursively placing higher resolution meshes,
`refinements`, in heavily clustered regions. The Virgo
simulations were done on two large Cray T3D parallel
supercomputers at the Edinburgh Parallel Computing Center and at the
Computing Center of the Max Plank Society in Garching. These
simulations are publicly available at
{\it http://www.mpa-garching.mpg.de/Virgo/virgoproject.html.}

In the simulations used here, the evolution of particles was
followed in the comoving box of size $L=239.5h^{-1}$ Mpc. The
number of particles was $N_p=256^3$. Therefore, the mean particle
separation $\lambda_p=L/N_p^{1/3}=0.9355 h^{-1}$ Mpc.
The mass of the particle
$m_p=\rho_b \lambda_p^3=6.82 \times 10^{10} h^{-1} M_{\odot}$ and
$m_p=2.27 \times 10^{11} h^{-1} M_{\odot}$ in the $\Lambda$CDM and
$\tau$CDM models, respectively (here $\rho_b$ is the mean
background density). The gravitational softening length is
$r_{soft}=25h^{-1}$ kpc and $r_{soft}=36h^{-1}$ kpc ,
respectively (see Jenkins et al. (1998) for a detailed
description of the force calculation scheme used in the Virgo
simulations). We denote the Virgo $\Lambda$CDM and $\tau$CDM models
as the model $\Lambda$CDM1 and $\tau$CDM.

We also investigated the evolution of $256^3$ particles on a
$256^3$ grid using a particle-mesh (PM) code described by Gramann
(1988) and Suhhonenko \& Gramann (1999). The PM code achieves the force
resolution close to the mean particle separation $\lambda_p$.
The PM method is discussed in detail by Hockney \& Eastwood (1981) and
Efstathiou et al.(1985). The cosmological parameters for this PM simulation
were chosen similar to the $\Lambda$CDM1 model. We chose
the flat $\Omega_0=0.3$ model and used the initial power spectrum
$P(k)$ given in eq. (1) (with $\sigma_8=0.9$). The comoving box
size was $L=239.5 h^{-1}$ Mpc. Therefore, the mean particle separation,
$\lambda_p$, and the mass of the particle,
$m_p$, in this model are the same as used in the $\Lambda$CDM1 model.
We denote this model as the $\Lambda$CDM2 model.

We examined the rms velocity of particles in the simulations
studied. In the $\Lambda$CDM1 and $\tau$CDM models, the rms
velocity of particles was $648 \kms$ and $636 \kms$, respectively.
In the $\Lambda$CDM2 model, the rms velocity was $575 \kms$. Due to the
small-scale smoothing inherent to the PM method, the intrinsic velocity
dispersions of clusters in the $\Lambda$CDM2 model are smaller than
in the $\Lambda$CDM1 model. The velocity field of clusters in
these models is expected to be similar.

In the $\Lambda$CDM2 model we used the traditional two-point
approximation to calculate the acceleration on the grid. We also
studied the evolution of particles using a PM code with the
shifted-mesh scheme. The effect of the shifted-mesh scheme on the
cluster velocities is discussed in Section 4.

\ssec{Selection of clusters}

We used two different algorithms to identify clusters in
simulations: the standard friends-of-friends (FOF) algorithm, and
the algorithm, where clusters are defined as maxima of the
density field smoothed on the scale $R\sim 1h^{-1}$ Mpc (DENSMAX).

The friends-of-friends group finder algorithm was applied using
the program suite developed by the cosmology group in the
University of Washington. These
programs are available at {\it http://www-hpcc.astro.washington.edu.}

The FOF cluster finder depends on one parameter $b$, which
defines the linking length as $b\lambda_p$. The conventional
choice for this parameter is $b=0.2$ (see e.g. G\"otz, Huchra \&
Brandenberger 1998; Jenkins et al. 2001). In this paper we also define
clusters by using the value $b=0.2$. We also study velocities
of the clusters defined by the parameters $b=0.15$ and $b=0.3$.
In the limit of very large numbers of particles per object, FOF
approximately selects the matter enclosed by an isodensity contour
at $1/b^3$.

\begin{figure}
\centering
\leavevmode
\psfig{file=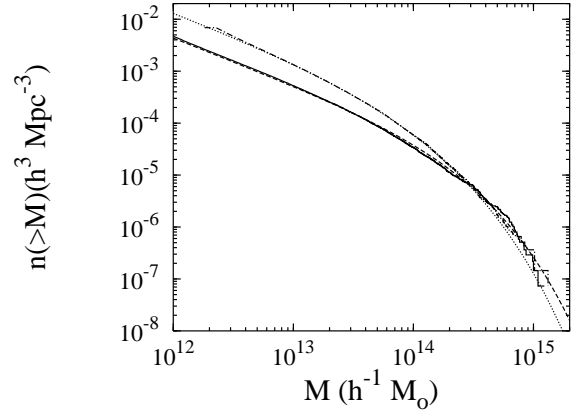,width=8cm}

\caption {The cluster mass function in the Virgo $\Lambda$CDM model
for the parameter $b=0.164$ (solid line) and in the $\tau$CDM
model for $b=0.2$ (dot-dashed line). For comparison we show the
fitting formulae given by Jenkins et al. (2001) for the mass
function in the $\Lambda$CDM (dashed line) and in the $\tau$CDM
model (dotted line).}

\end{figure}

To test our FOF output data, we found the mass function of
clusters in the Virgo simulations. The cluster mass function in
these simulations has been studied in detail by Jenkins et al. (2001).
Fig. 2 shows the mass function of clusters determined for
$b=0.164$ in the $\Lambda$CDM model and for $b=0.2$ in the
$\tau$CDM model. For comparison we show the mass functions
given by the approximations obtained by
Jenkins et al. (2001) (eq. (B2) and (B1) for the $\Lambda$CDM and
$\tau$CDM model, respectively). Jenkins et al. (2001) studied the
mass function at the high mass end up to the point where the
predicted Poisson abundance errors reach $10$\%. In the
simulations studied here, this limit is reached when the number
density of clusters is $n=7.3 \times
10^{-6} \mpc3h$. Fig. 2 shows that if $n$ is larger than this
value, the agreement between our results and these obtained by
Jenkins et al. (2001) is very good.

We studied clusters that contained at least ten particles.
The three-dimensional velocity of each cluster was defined as
$$
\vec v_{cl} = {1\over N_p} \sum_{i=1}^{N_{p}} \vec v_i, \eqno(5)
$$
where $N_{p}$ is the number of particles in the cluster and $\vec
v_i$ is the velocity of the particle $i$ in the cluster. To
characterize the size of the cluster, we use the effective radius
defined as
$$
R_{eff} = {1\over N_{p}}{\sum_{i=1}^{N_{p}}{\bigl[(x_{i}-\bar
x)^2+(y_{i}-\bar y)^2+ (z_{i}-\bar z)^2}\bigr]^{1/2}},
\eqno(6)
$$
where $x_i,y_i,z_i$ are the particle coordinates in the cluster
and ($\bar x, \bar y, \bar z$) are the coordinates of the cluster
centre. If we use the FOF method, the mean size for the high-mass
clusters is larger than the mean size for the low-mass clusters.

We also selected clusters using the DENSMAX method. In this
case clusters were identified in the simulations as maxima of
the density field that was determined on a $256^3$ grid using the
cloud-in-cell (CIC) scheme. To determine peculiar velocities of
clusters, $v_{cl}$, we calculated the peculiar velocity field on
a $256^3$ grid using the CIC-scheme, and found peculiar
velocities at the grid points where the clusters had been identified.
If we use the DENSMAX method, the size of the selected clusters is
similar for all clusters, given by the cell size, $\lambda_g$.
In our case, the cell size $\lambda_g=0.9355h^{-1}$ Mpc.

To determine the rms peculiar velocities of clusters, we used the
equation
$$
v_{rms}^2= v_s^2 + v_L^2 = {1\over N_{cl}}{\sum_{i=1}^{N_{cl}}
v_{cli}^2} +v_L^2, \eqno (7)
$$
where the parameter $v_s$ describes the dispersion of cluster
velocities, $v_{cli}$, derived from simulation and the parameter
$v_L$ is the linear contribution from velocity fluctuations on scales
greater than the size of the simulation box $L$ and is given by
$$
v_L^2 ={f^2(\Omega_0)H_0^2 \over{2\pi^2}}\int_{0}^{2\pi \over
L}P(k)dk. \eqno (8)
$$
$N_{cl}$ is the number of clusters studied. By using eq. (4),
the linear rms peculiar velocity of peaks can be written as
$$
\sigma_p^2(R)=\sigma_v^2(R) - H_0^2 f^2(\Omega_0) {\sigma_0^4(R)
\over \sigma_1^2(R)}  . \eqno(9)
$$
The second term in this expression is not sensitive to the
amplitude of large-scale fluctuations at wavenumbers $k<2\pi/L$.
Therefore, the linear rms velocity of peaks can be expressed as
$$
\sigma_p^2(R) = \sigma_p^{\prime 2}(R) +v_L^2, \eqno(10)
$$
where $\sigma_p^{\prime} (R)$ is determined by the power
spectrum at wavenumbers $k>2\pi/L$ and $v_L$ is given by eq. (8).
For $\Lambda$CDM and $\tau$CDM models, we found that $v_L = 220
\kms$ and $v_L = 245 \kms$, respectively.

If the one-dimensional velocities of clusters, $v_{xi}$, follow a
Gaussian distribution with a mean $\bar v_{x}=0$ and a dispersion
$\sigma^2$, then the sum
$$
\chi^2={1 \over \sigma^2} {\sum_{i=1}^{N_{cl}} v_{cli}^2}
\eqno(11)
$$
is distributed as a $\chi^2$ distribution with the number of
degrees of freedom $\nu=3 N_{cl}$. In this case, the rms error
for the variable $v_{rms}^2$ can be determined as
$$
\Delta v_{rms}^2 = \sqrt {2 \over 3 N_{cl}} \,\, v_s^2. \eqno(12)
$$
As a first step, the one-dimensional distribution of cluster
velocities can be approximated as a Gaussian distribution (see
e.g. Bahcall, Gramann, Cen (1994) for the study of the velocity
distribution of clusters in different cosmological models). We
used eq. (12) to estimate the error bars for the rms velocities
of clusters.

\sec{RESULTS}

\begin{table}
\caption{The number of clusters, $N_{cl}$, in different density and mass
intervals. $N_{cl}$ for different density
intervals is given for the clusters selected with the DENSMAX
method. $N_{cl}$ for different mass intervals is
given for the clusters determined with the FOF method with $b=0.2$.}

\begin{tabular}{|c|r|r|r|}
\hline \hline
 $\rho/\rho_b$    & $N_{cl}$          & $N_{cl}$  & $N_{cl}$  \\
           & $\Lambda$CDM1   & $\tau$CDM& $\Lambda$CDM2     \\
\hline
            1 --    5 & 148496 & 169774 & 103757  \\
            5 --   10 & 29265 & 35753 & 17369  \\
           10 --   50 & 31294 & 34705 & 18314  \\
           50 --  100 & 5246 & 4015 & 3372  \\
          100 --  500 & 4253 & 1762 & 2619  \\
          500 -- 1000 & 298 & 22 & 107  \\
         1000 -- 5000 & 78 &  0 &  8 \\
\hline \hline
 $M$ & $N_{cl}$          & $N_{cl}$  &$N_{cl}$  \\
   ($h^{-1}M_{\odot}$)  &$\Lambda$CDM1 & $\tau$CDM & $\Lambda$CDM2 \\
\hline
           $5\times 10^{11}$ -- $10^{12}$ & 36227 & 0 & 27310  \\
           $10^{12}$ -- $5\times 10^{12}$ & 51841 & 57460 & 15820  \\
           $5\times 10^{12}$ -- $10^{13}$ & 7326 & 19298 & 3204  \\
           $10^{13}$ -- $5\times 10^{13}$ & 6542 & 15489 & 4225  \\
           $5\times 10^{13}$ -- $10^{14}$ & 856 & 1644 & 659  \\
           $10^{14}$ -- $5\times 10^{14}$ & 511 & 789 & 507  \\
           $5\times 10^{14}$ -- $10^{15}$ & 38 & 18 & 24  \\
         $10^{15}$ -- $5\times 10^{15}$ & 4 & 5 & 3  \\
\hline
\end{tabular}
\end{table}

\begin{figure*}
\centering \leavevmode \psfig{file=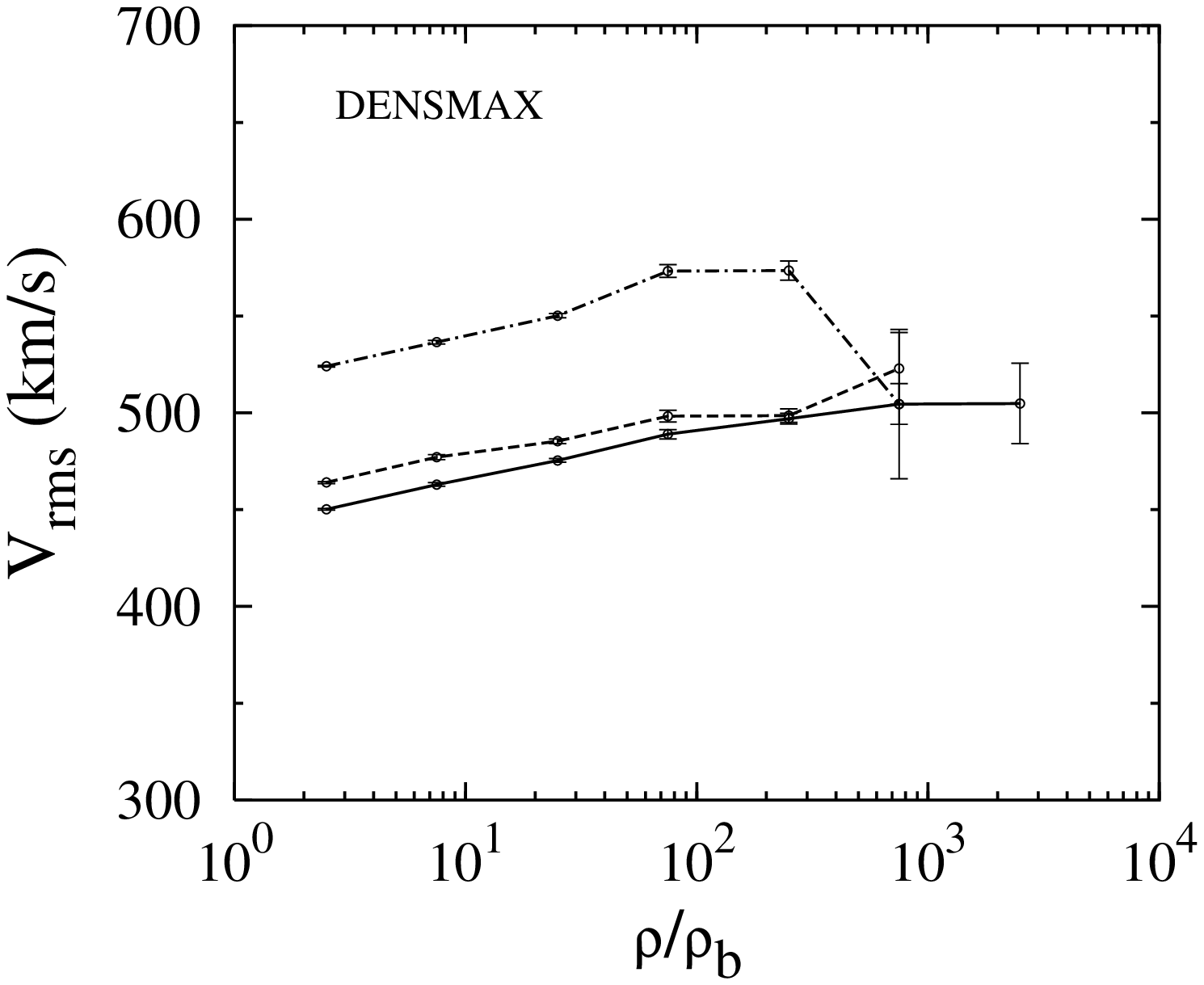,width=8cm}
\psfig{file=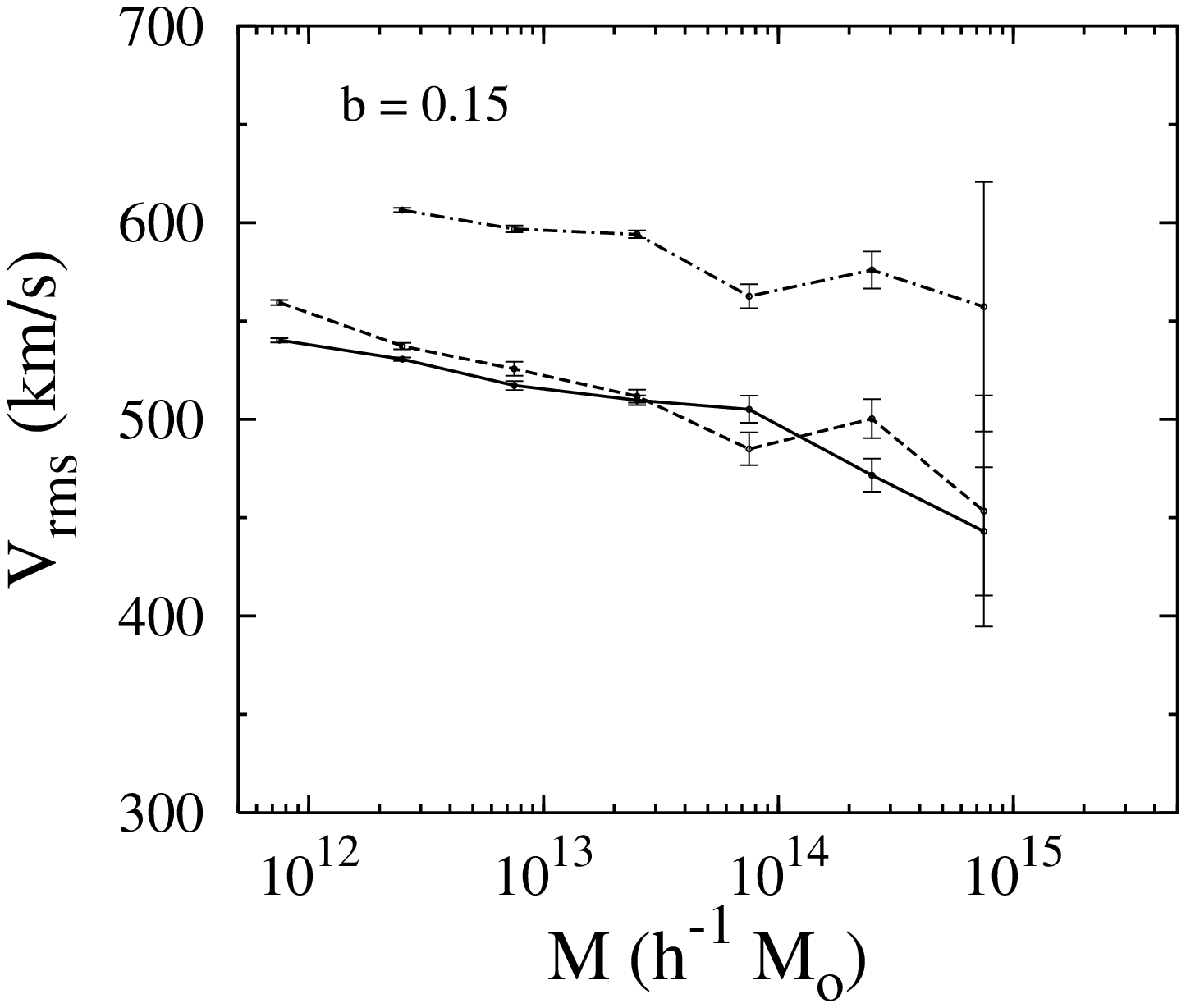,width=8cm} \psfig{file=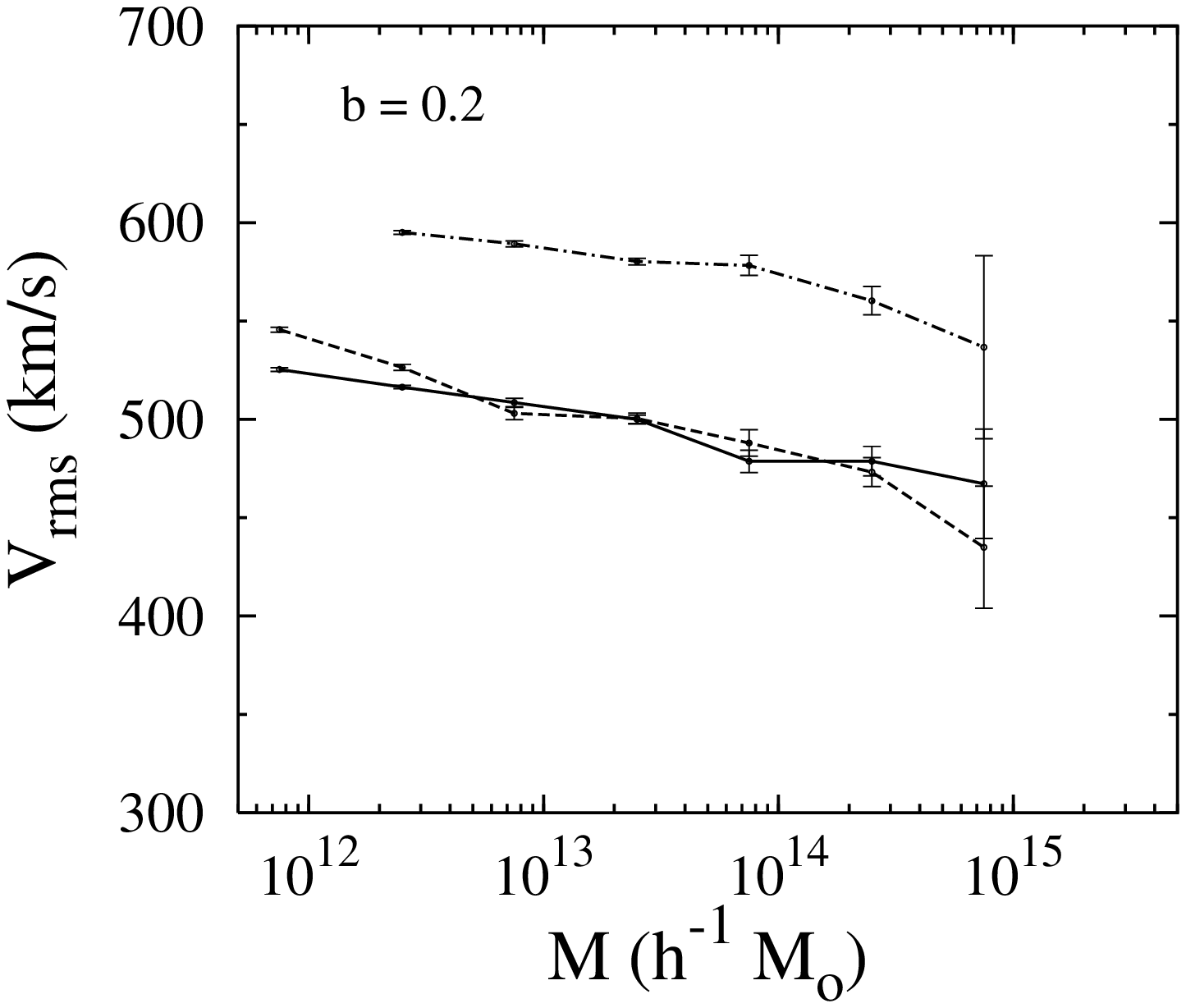,width=8cm}
\psfig{file=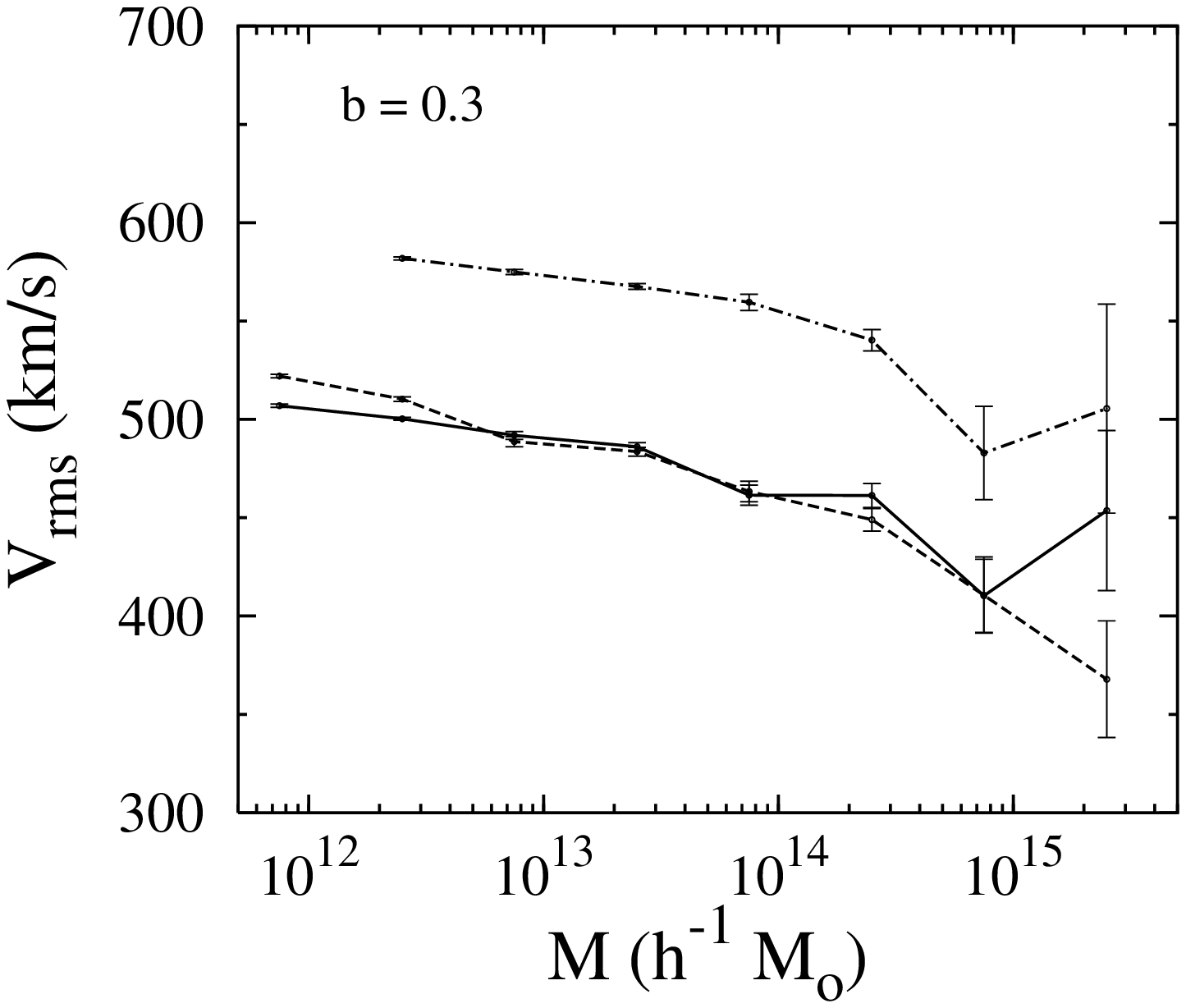,width=8cm}

\caption{The rms peculiar velocities of clusters for different
densities and masses in the $\Lambda$CDM1 model (solid lines),
$\tau$CDM model (dot-dashed lines) and in the $\Lambda$CDM2 model
(dashed lines). The upper left panel shows the results for the
clusters defined by the DENSMAX method and the lower left panel
shows the results for the clusters selected by the FOF method
with $b=0.2$. The right panels show the rms velocities of
clusters defined by the FOF method with $b=0.15$ and $b=0.3$,
respectively.}

\end{figure*}

First, we investigated the rms velocity of clusters in different
density and mass intervals. The results are presented in Fig. 3.
The rms velocities are shown for the intervals, where the number
of clusters is $N_{cl}>10$.

The clusters defined with the DENSMAX method were divided into
subgroups according to their density. We studied the rms velocity
of clusters in seven subgroups, where the density $\rho/\rho_b$
was in the range $1 - 5$, $5 - 10$, $10 - 50$,...,$1000 - 5000$.
Table~1 shows the number of clusters and the upper right panel in
Fig.~3 shows the rms velocity of clusters in different density
intervals. We see that the rms velocities of clusters in the
$\Lambda$CDM1 and $\Lambda$CDM2 models are similar. In the models
studied, the rms velocity of clusters is almost independent of
the density. The rms velocity somewhat increases at smaller
densities. However, this increase is small ($\approx 10$ per
cent). For the range $\rho/\rho_b = 100-500$, the rms velocity is
$505$ km s$^{-1}$ and $570$ km s$^{-1}$ in the $\Lambda$CDM1 and
$\tau$CDM models, respectively. These values are similar to the
linear theory expectations. At the radius $R=1h^{-1}$ Mpc, the
rms peculiar velocity of peaks $\sigma_p=509$ km s$^{-1}$ and
$562$ km s$^{-1}$ in the $\Lambda$CDM1 and $\tau$CDM models,
respectively. The rms velocity for the low-density clusters is
somewhat smaller than predicted by the linear theory.

The clusters determined with the FOF method were divided into subgroups
according to their mass. We studied the rms velocity of clusters
in  eight subgroups, where the mass was in the range $(5 \times 10^{11}
- 10^{12})h^{-1}M_{\odot}$, ..., $(10^{15} - 5\times 10^{15}) h^{-1}
M_{\odot}$. Table~3 shows the number of clusters and Fig.~3 demonstrates
the rms peculiar velocity of clusters in different mass intervals.
The lower left panel in Fig.~3 shows the results for the clusters determined
by $b=0.2$. The right panels show the rms velocities of clusters
defined by $b=0.15$ and $b=0.3$.

We see that in the $\Lambda$CDM1 and $\Lambda$CDM2 models the rms
velocities are similar. The rms velocity of FOF clusters
decreases with cluster mass. The rms velocity of massive FOF clusters
is smaller than the rms velocity of high-density clusters determined
with the DENSMAX method. In the $\Lambda$CDM model for $b=0.2$,
the rms velocity of clusters is $525 \kms$ in the mass interval
$(5\times 10^{11} -10^{12})h^{-1} M_{\odot}$
and $430 \kms$ in the mass interval
$(10^{15} - 5\times 10^{15})h^{-1} M_{\odot}$. For $b=0.15$ and $b=0.3$,
this effect is similar. These results are in good agreement with the results
obtained by Sheth \& Diaferio (2001). They studied the rms velocities of
clusters in different mass intervals and found that the rms
cluster velocity decreases with mass.

\begin{figure*}
\centering \leavevmode \psfig{file=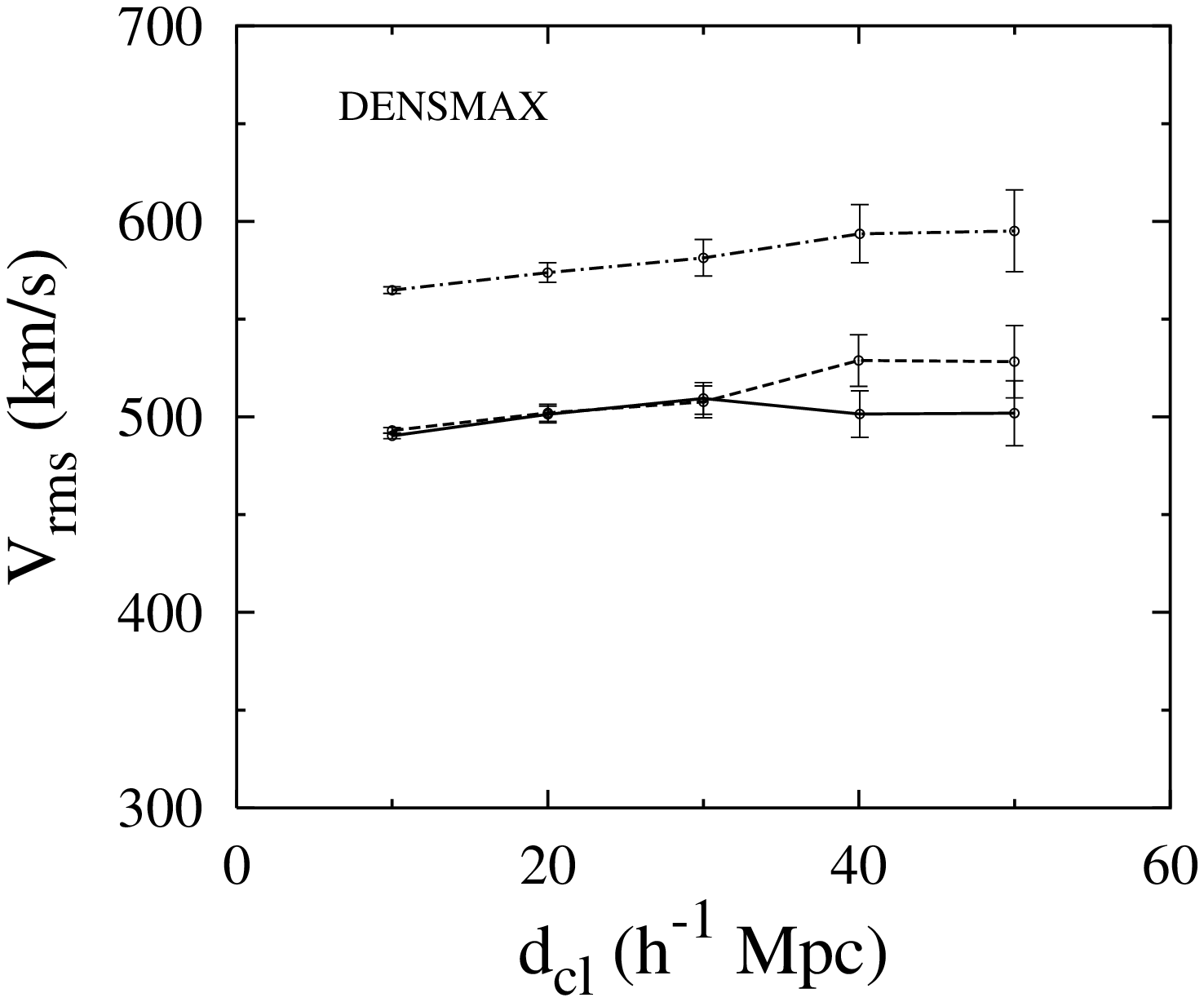,width=8cm}
\psfig{file=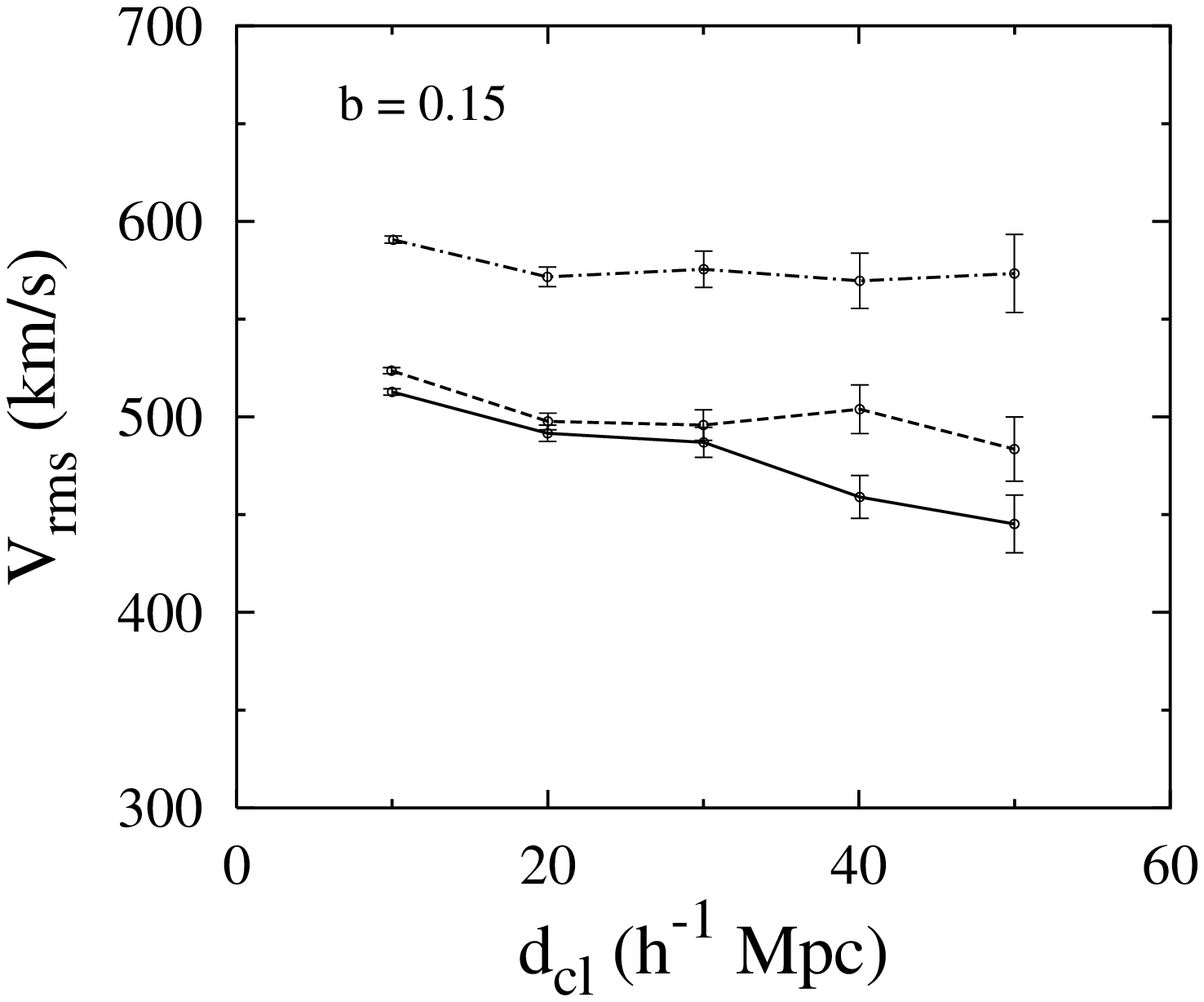,width=8cm} \psfig{file=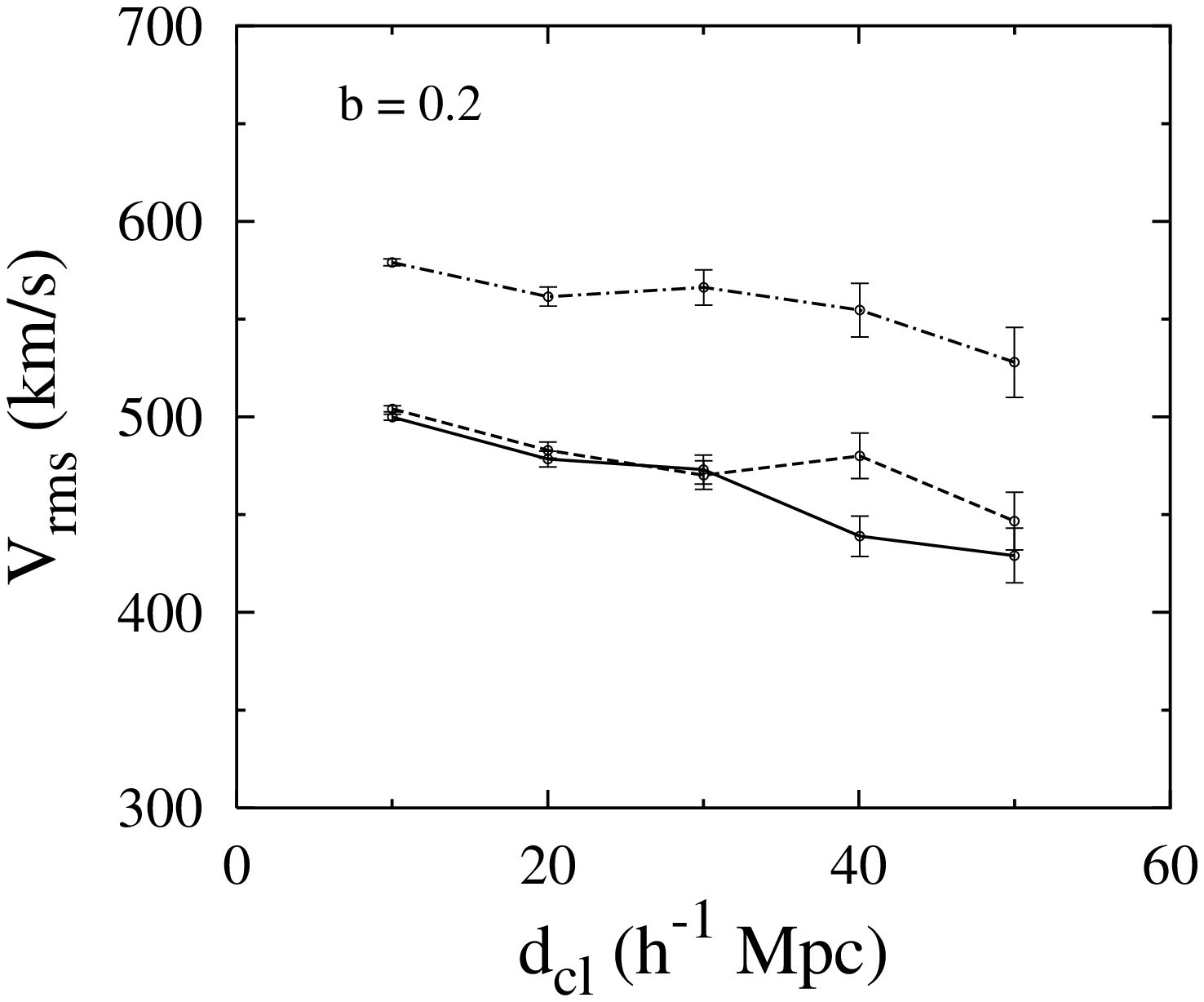,width=8cm}
\psfig{file=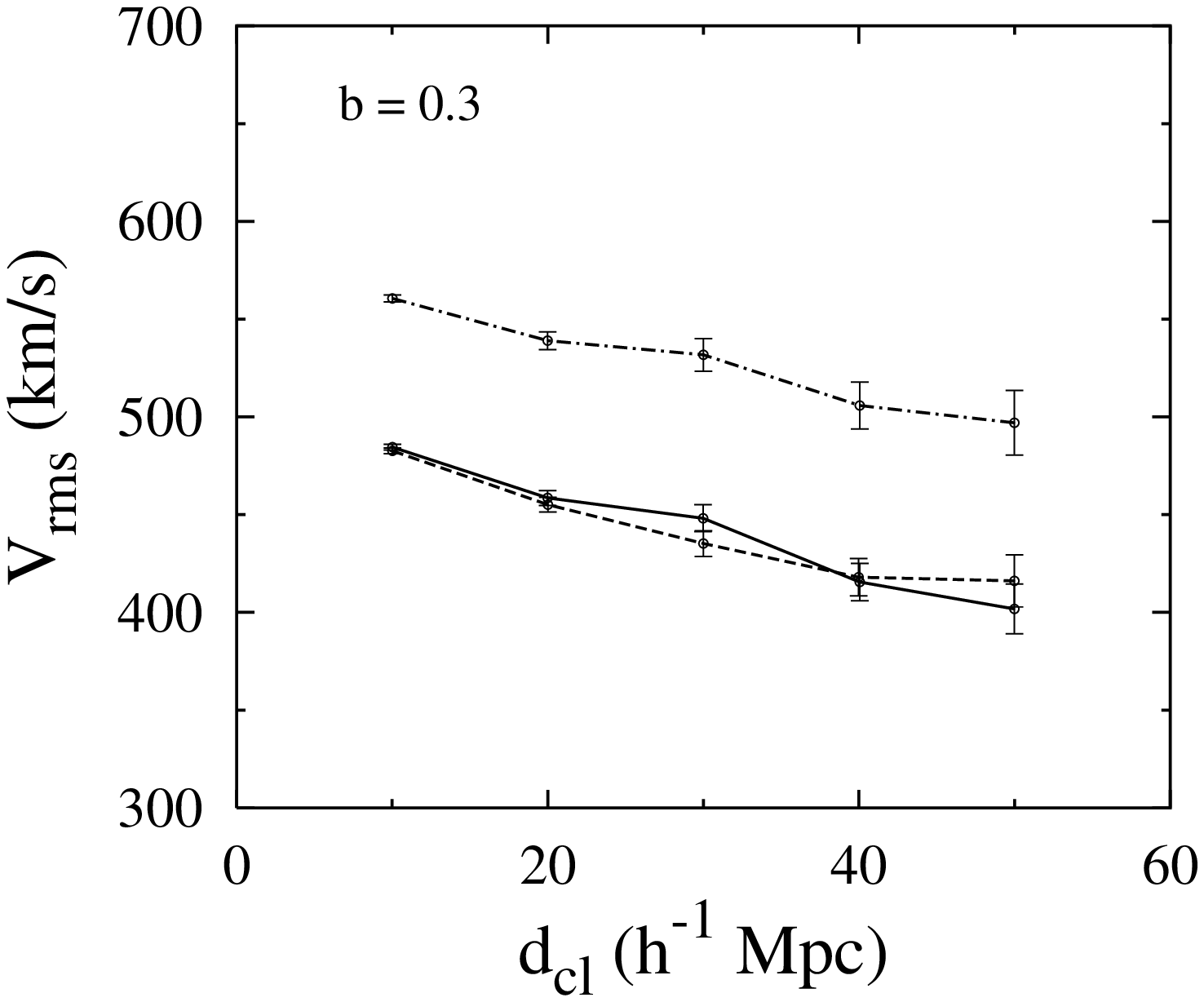,width=8cm}

\caption{The rms peculiar velocities of clusters for different
values of the mean cluster separation, $d_{cl}$. The clusters are
ranked according to their density (DENSMAX clusters) or mass (FOF
clusters). The lines are defined as in Fig.~3.}

\end{figure*}

\begin{table}

\caption{The density and mass thresholds used to produce cluster
catalogues with mean separations $d_{cl}=10-50h^{-1}$ Mpc. The
mass thresholds are given for the FOF clusters determined by
$b=0.2$.}
\begin{tabular}{|c|c|c|c|}
\hline
\hline
 $d_{cl}$ & $\rho_{t}/\rho_b$ & $\rho_{t}/\rho_b$ & $\rho_{t}/\rho_b$  \\
   ($\mpcoh$)  &$\Lambda$CDM1  & $\tau$CDM & $\Lambda$CDM2 \\
\hline
           10 & 35  & 27 & 20  \\
           20 & 213 & 102 & 136  \\
           30 & 435 & 178 & 264  \\
           40 & 652 & 254 & 393  \\
           50 & 850 & 315 & 505  \\
\hline
\hline
 $d_{cl}$ & $M_t$ ($\Lambda$CDM1)        & $M_t$ ($\tau$CDM) &
 $M_t$ ($\Lambda$CDM2) \\
   ($\mpcoh$)  & ($h^{-1}M_{\odot}$)& ($h^{-1}M_{\odot}$) &
   ($h^{-1}M_{\odot}$) \\
\hline
     10 & $5.6 \times 10^{12}$ & $1.3 \times 10^{13}$& $2.3 \times 10^{12}$  \\
     20 & $4.3 \times 10^{13}$ & $6.4 \times 10^{13}$& $3.6 \times 10^{13}$  \\
     30 & $1.1 \times 10^{14}$ & $1.3 \times 10^{14}$& $1.1 \times 10^{14}$  \\
     40 & $2.0 \times 10^{14}$ & $2.0 \times 10^{14}$& $1.9 \times 10^{14}$  \\
     50 & $3.0 \times 10^{14}$ & $2.8 \times 10^{14}$& $2.8 \times 10^{14}$  \\
\hline
\end{tabular}
\end{table}

We also studied the rms velocity of clusters with density (or
mass) higher than a given threshold density (or mass). The
results are presented in Fig.~4. The DENSMAX clusters were ranked
according to their density, and we selected $N_{cl}=(L/d_{cl})^3$
highest ranked clusters to produce cluster catalogues with a mean
intercluster separation $10-50h^{-1}$ Mpc. Similarly, the FOF
clusters were ranked according to their mass. Table~2 shows the
density and mass thresholds used to produce cluster
catalogues for different values of the mean cluster separation.
For comparison, the number density of observed APM clusters and
Abell clusters is $n_{cl}\sim 3.4 \times 10^{-5}h^{3}$ Mpc$^{-3}$
($d_{cl}\sim 31 h^{-1}$Mpc) and $n_{cl}\sim 2.5 \times
10^{-5}h^{3}$ Mpc$^{-3}$ ($d_{cl}\sim 34h^{-1}$ Mpc),
respectively (Dalton et al. 1994, Einasto et al. 1997).

Fig.~4 demonstrates the rms peculiar velocity of clusters with a
mean separation $d_{cl}=10 - 50h^{-1}$ Mpc. In this range, the rms
velocity of DENSMAX clusters is almost independent of the density of
clusters. For the clusters with a mean separation $d_{cl}=30h^{-1}$ Mpc,
$v_{rms}=510 \kms$ and $v_{rms}=580 \kms$ in the $\Lambda$CDM1 and
$\tau$CDM models, respectively. These values are similar to the
linear rms velocity of peaks at the radius $R\sim 1h^{-1}$Mpc.

The rms velocity of FOF clusters decreases with cluster richness.
For rich clusters, the rms velocity of FOF clusters is smaller
than the rms velocity of clusters determined with the DENSMAX
method. In the $\Lambda$CDM model for $b=0.2$, the rms velocities
are $500 \kms$ and $430 \kms$, if $d_{cl}=10h^{-1}$ Mpc and
$d_{cl}=50h^{-1}$ Mpc, respectively. For $b=0.15$ and $b=0.3$,
this effect is similar. For the clusters with $d_{cl}=30h^{-1}$
and $b=0.2$, $v_{rms}=475 \kms$ and $v_{rms}=565 \kms$ in the
$\Lambda$CDM1 and $\tau$CDM models, respectively.

In Table~3 we compare the rms velocity of FOF clusters,
$v_{rms}$, with $\sigma_p(R)$ for the radius $R=1h^{-1}$Mpc. We
analyzed the rms velocity of clusters for different values of the
cluster separation. The results are given for the clusters
determined by $b=0.2$. The rms peculiar velocities $v_{rms}$ for
the clusters determined by $b=0.15$ and $b=0.3$ are similar. In
the $\Lambda$CDM1 model, the rms peculiar velocity of small
clusters is close to the linear theory expectations, while the
rms peculiar velocity of rich clusters is smaller ($\approx$15\%
per cent for clusters with a mean intercluster separation
$d_{cl}=50h^{-1}$ Mpc).

\begin{table}
\caption{Comparison of the rms velocity of FOF clusters, $v_{rms}$, with
the linear theory predictions for peculiar velocities of peaks, $\sigma_p$,
for the radius $R=1h^{-1}$ Mpc. The results are given for the FOF clusters
determined by $b=0.2$.}

\begin{tabular}{|c|c|c|c|}
\hline \hline
 $d_{cl}$        & $v_{rms}/\sigma_p$  & $v_{rms}/\sigma_p$   &
 $v_{rms}/\sigma_p$   \\
 ($\mpcoh$)      & $\Lambda$CDM1      & $\tau$CDM       & $\Lambda$CDM2 \\
\hline
           10 & 0.98 & 1.03 & 0.99  \\
           30 & 0.93 & 1.01 & 0.92  \\
           50 & 0.84 & 0.94 & 0.88  \\
\hline
\end{tabular}
\end{table}

Our results are in good agreement with the results obtained by
Colberg et al. (2000). They also studied the rms velocity of
clusters in the Virgo $\Lambda$CDM and $\tau$CDM models, but used a
slightly different method to select clusters. High-density
regions were located using a FOF method with $b=0.05$ and their
barycentres were considered as candidate cluster centers. Any
candidate centre for which mass within $1.5h^{-1}$ Mpc exceeded the
threshold mass $M_t$ was identified as a candidate cluster. The final cluster
list was obtained by deleting the lower mass candidate in all
pairs separated by less than $1.5h^{-1}$ Mpc. The peculiar
velocity of each cluster was defined to be the mean peculiar
velocity of all the particles within the $1.5h^{-1}$ Mpc sphere.
In this method, the size of the selected clusters is same for all
clusters and in this sense, this method is similar to the DENSMAX
method.

Colberg et al. (2000) used the value $M_t=3.5 \times 10^{14}
h^{-1} M_{\odot}$. For this value, the number of clusters was
$\approx 70$ in the $\Lambda$CDM1 and $\tau$CDM models ($d_{cl} \approx
58h^{-1}$Mpc). They found that the rms cluster velocities derived
from simulation, $v_s$ are $439 \kms$ and $535 \kms$ in the
$\lambda$CDM1 and $\tau$CDM models, respectively (they did not include the
dispersion $v_{L}^2$). If we use the DENSMAX method, we find that
the rms velocity of clusters is almost independent of the number density of
clusters and for $d_{cl}=50h^{-1}$ Mpc,
the velocities are $v_s=450 \kms$ and $549 \kms$ in the $\Lambda$CDM1 and
$\tau$CDM models, respectively. These values are very close to the
values found by Colberg et al. (2000), only slightly larger
($\approx 2$ per cent). This small difference is probably caused by
the fact that in the DENSMAX method we use the smoothing length
$\sim 1h^{-1}$ Mpc, which is smaller than $1.5h^{-1}$Mpc used by
Colberg et al. (2000)]. For
comparison, the rms velocities $v_s$ for the FOF clusters for $b=0.2$
and $d_{cl}=50h^{-1}$ Mpc, are $368 \kms$ and $468 \kms$ in the
$\Lambda$CDM and $\tau$CDM models, respectively.

Let us now consider the rms velocities of clusters for different
cluster radii. We studied the rms velocity of clusters with
the effective radius $R_{eff}$ larger than a given threshold
radius. The clusters were ranked according to their effective
radius and we
selected $N_{cl}=(L/d_{cl})^3$ highest ranked clusters to produce
cluster catalogues with mean separations $10-50h^{-1}$ Mpc.
Table~4 shows the threshold radii used for different values of
the mean cluster separation.

\begin{table}
\caption{The threshold radii $R_t$ used to produce
cluster catalogues with mean separations $d_{cl}=10 - 50h^{-1}$ Mpc. The
radii are given for the FOF clusters defined by $b=0.2$.}
\begin{tabular}{|c|c|c|}
\hline \hline
 $d_{cl}$        & $R_t$ ($\Lambda$CDM1) & $R_t$ ($\tau$CDM)    \\
 ($\mpcoh$)      & ($\mpcoh$)         &($\mpcoh$)              \\
\hline
           10 & 0.22 & 0.24  \\
           20 & 0.40 & 0.40  \\
           30 & 0.58 & 0.53  \\
           40 & 0.74 & 0.64  \\
           50 & 0.88 & 0.72  \\
\hline
\end{tabular}
\end{table}

\begin{figure}
\centering \leavevmode \psfig{file=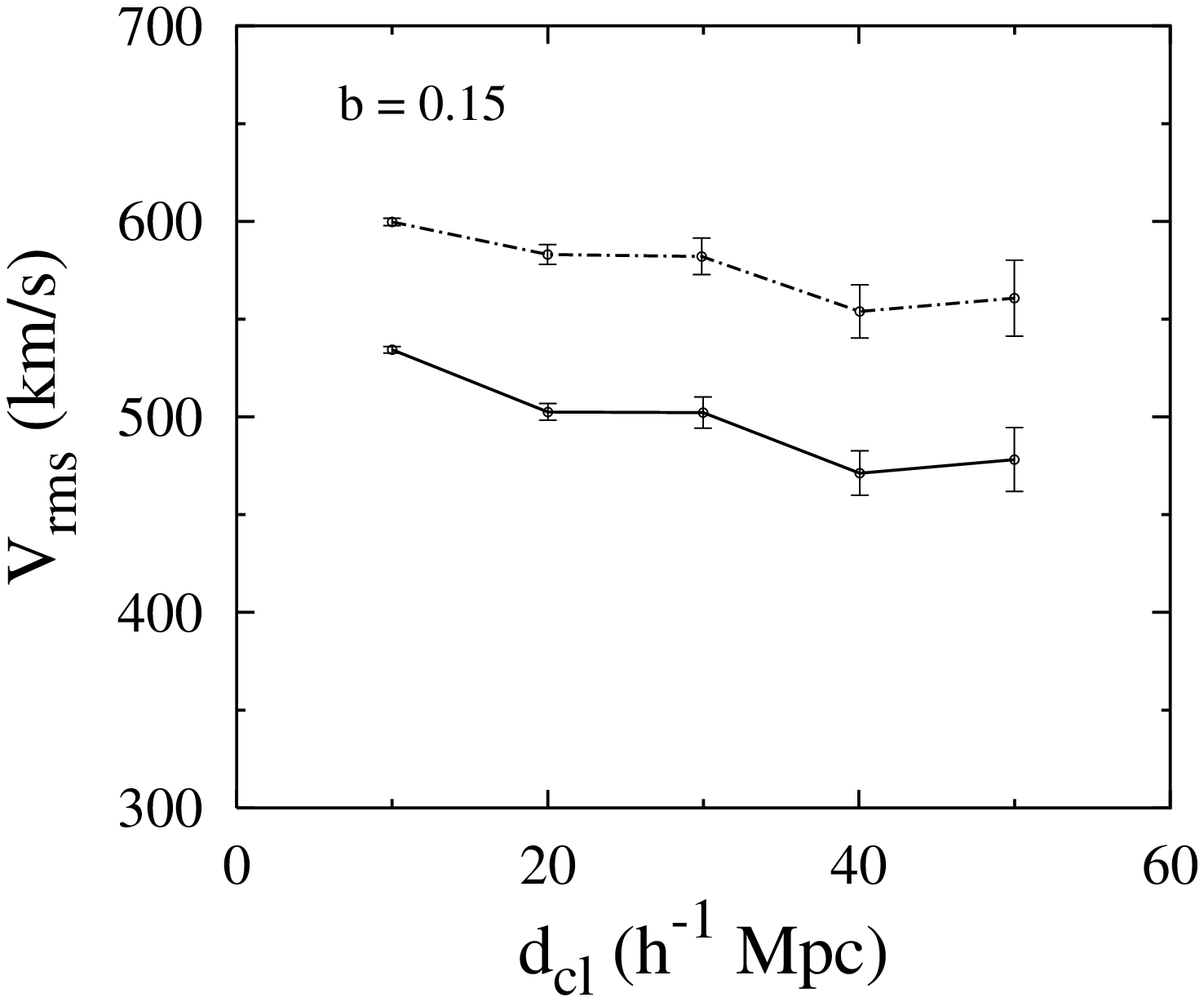,width=8cm}
\psfig{file=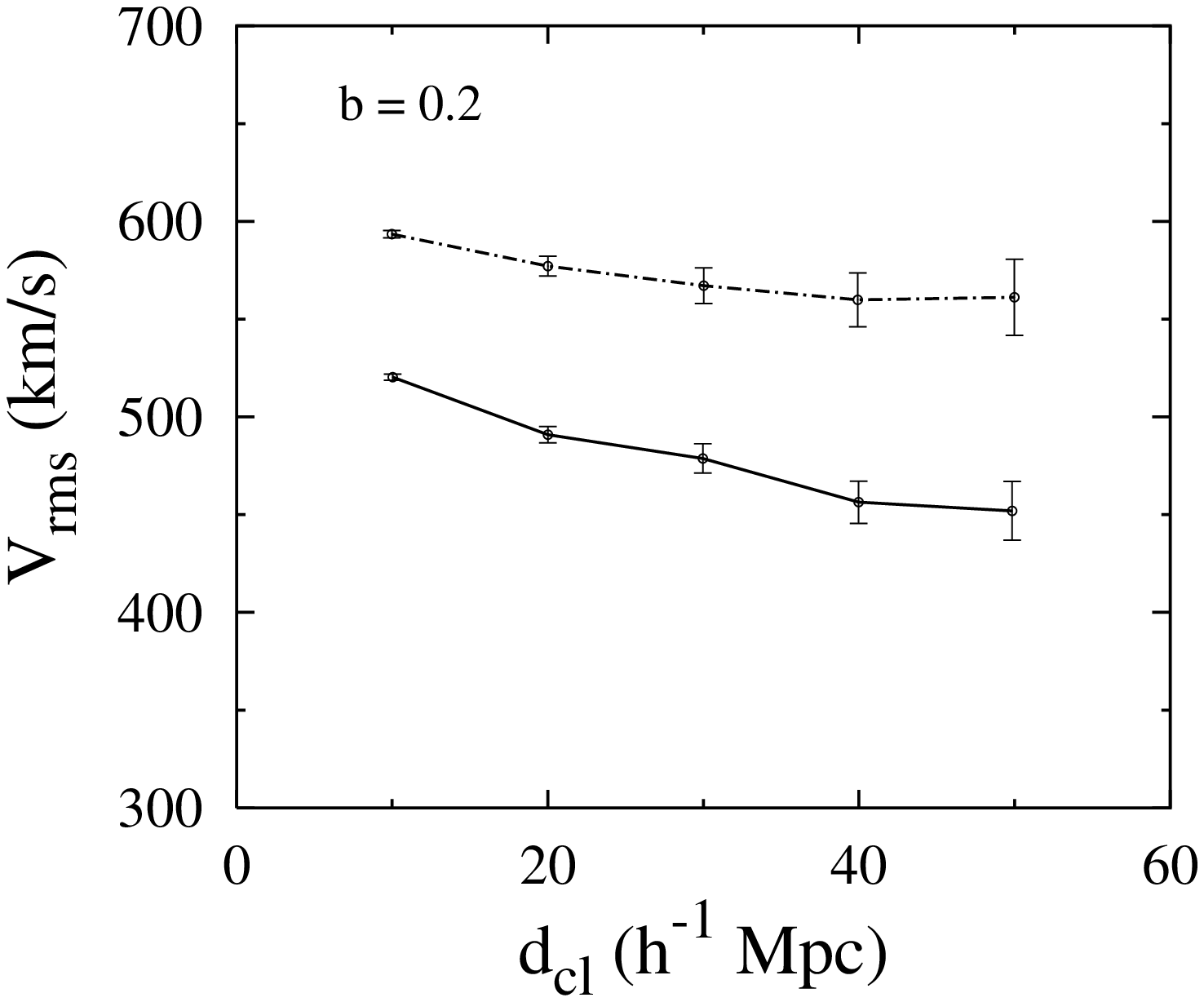,width=8cm} \psfig{file=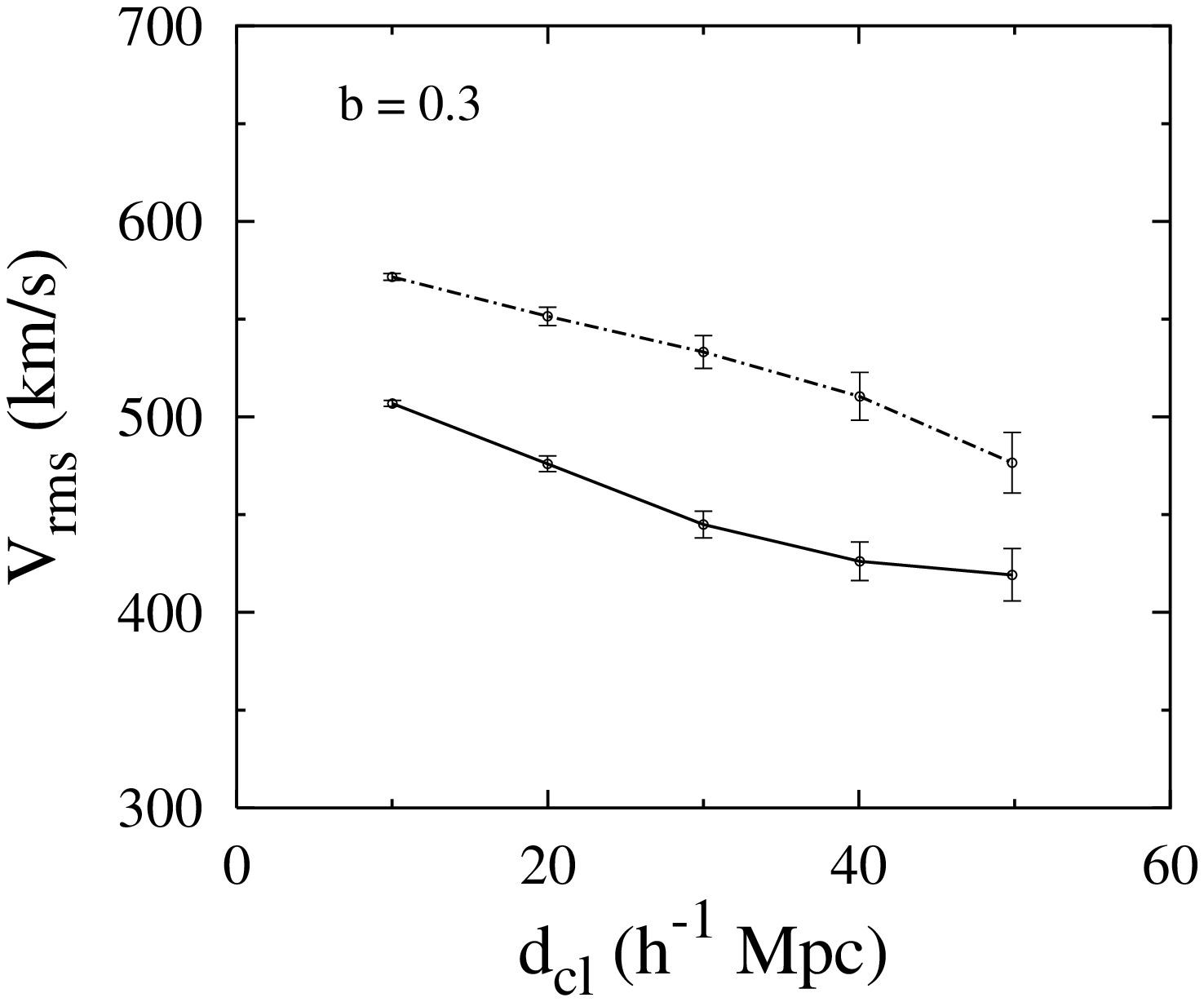,width=8cm}

\caption{The rms peculiar velocities of clusters for different
values of the mean cluster separation, $d_{cl}$. The clusters are
ranked according to their effective radius, $R_{eff}$. The solid lines
show the rms velocities in the $\Lambda$CDM1 model and the dot-dashed
lines in the $\tau$CDM model. The upper panel shows the
velocities for the clusters defined by $b=0.15$, the middle
panel for the clusters defined by $b=0.2$ and lower panel for
the clusters defined by $b=0.3$.}

\end{figure}

Fig. 5 illustrates the dependence of $v_{rms}$ on cluster
radii. We see that the effect of the cluster radius on $v_{rms}$ is
similar to the effect of the cluster mass on $v_{rms}$ (compare Fig.~4
and Fig.~5). In the models studied, the rms
velocity of small clusters is higher than the rms
velocity of large clusters. For example, in the $\Lambda$CDM1
model for $b=0.2$, the rms velocities are $520 \kms$ and $450 \kms$
for clusters with $d_{cl}=10h^{-1}$ Mpc and $d_{cl}=50h^{-1}$
Mpc, respectively.

\sec{THE SHIFTED-MESH SCHEME}

When we started the study of cluster velocities we first used a
PM code with a shifted-mesh scheme. In this scheme, the
acceleration on the grid was calculated by using the approximation
$$
g_x(i+{1\over 2},j,k)=\varphi(i,j,k)-\varphi(i+1,j,k),
\eqno(13)
$$
where $\varphi(i,j,k)$ is the gravitational potential on the grid.
A similar scheme for one-dimensional systems was proposed by Melott
(1986). The traditional two-point finite-difference approximation for
the acceleration is given by
$$
g_x(i,j,k)={\varphi(i-1,j,k) - \varphi(i+1,j,k) \over 2} .
\eqno(14)
$$
Gramann (1987) studied the evolution of an one-dimensional
plane-wave perturbation by using a PM code with differential
operators (13) and (14) and found that the approximation (13)
leads to smaller deviations from the exact solution than the
approximation (14).

Fig.~6 shows the rms peculiar velocities of clusters in a PM
simulation, where we used the approximation (13) for a
three-dimensional system. The cosmological parameters in this
simulation were chosen similar to the $\Lambda$CDM1 and
$\Lambda$CDM2 models. We investigated the evolution of $256^3$
particles on a $256^3$ grid in the comoving box of size
$L=239.5h^{-1}$Mpc. For comparison we show the rms peculiar
velocities in the $\Lambda$CDM2 model with a standard
finite-difference approximation (14). The clusters are defined by
the FOF method with $b=0.2$. The rms cluster velocities in the
$\Lambda$CDM2 model are similar to the $\Lambda$CDM1 model, which
achieves a force resolution smaller than the mean particle
separation $\lambda_p$ (see Fig.~3 and Fig.~4).

\begin{figure}
\centering \leavevmode \psfig{file=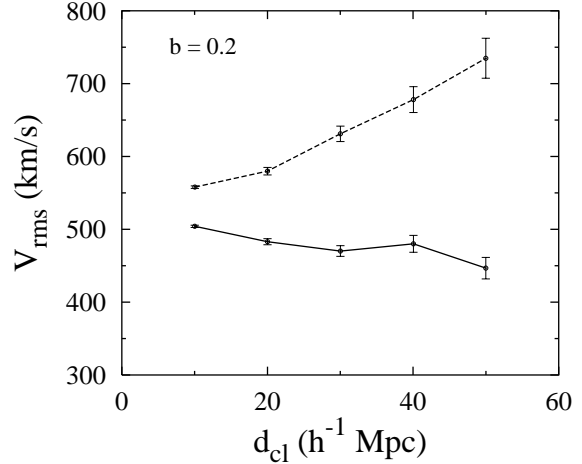,width=8cm}

\caption {The rms peculiar velocities of clusters in the PM
simulation, where we used the shifted-mesh scheme (13) (dashed
line) and in the $\Lambda$CDM2 model using a standard approximation
(14) (solid line)}.
\end{figure}

We see that the shifted-mesh scheme artificially boosts
cluster velocities. The effect is particularly strong for rich
clusters. The rms velocities are $560$ km s$^{-1}$ and $735$
kms$^{-1}$, if $d_{cl}=10h^{-1}$ Mpc and $d_{cl}=50h^{-1}$ Mpc,
respectively. This effect is probably caused by the fact that
using the scheme (13), we calculate the components ($g_x,g_y,g_z$)
at different locations on the grid and this can lead to
artificial enhancement of particle acceleration in one dimension.
This effect does not arise in an one-dimensional system, where
$g_y=g_z=0$.

In the $\Lambda$CDM2 model we used the standard two-point
finite-difference approximation (14). In this model, the rms
velocities of clusters are similar to the $\Lambda$CDM1 model.
Therefore, we can use the PM code to study peculiar velocities on
scales that are close to the mean particle separation and force
resolution. But it is important to use a correct difference
operator in simulations.

\begin{figure}
\centering \leavevmode \psfig{file=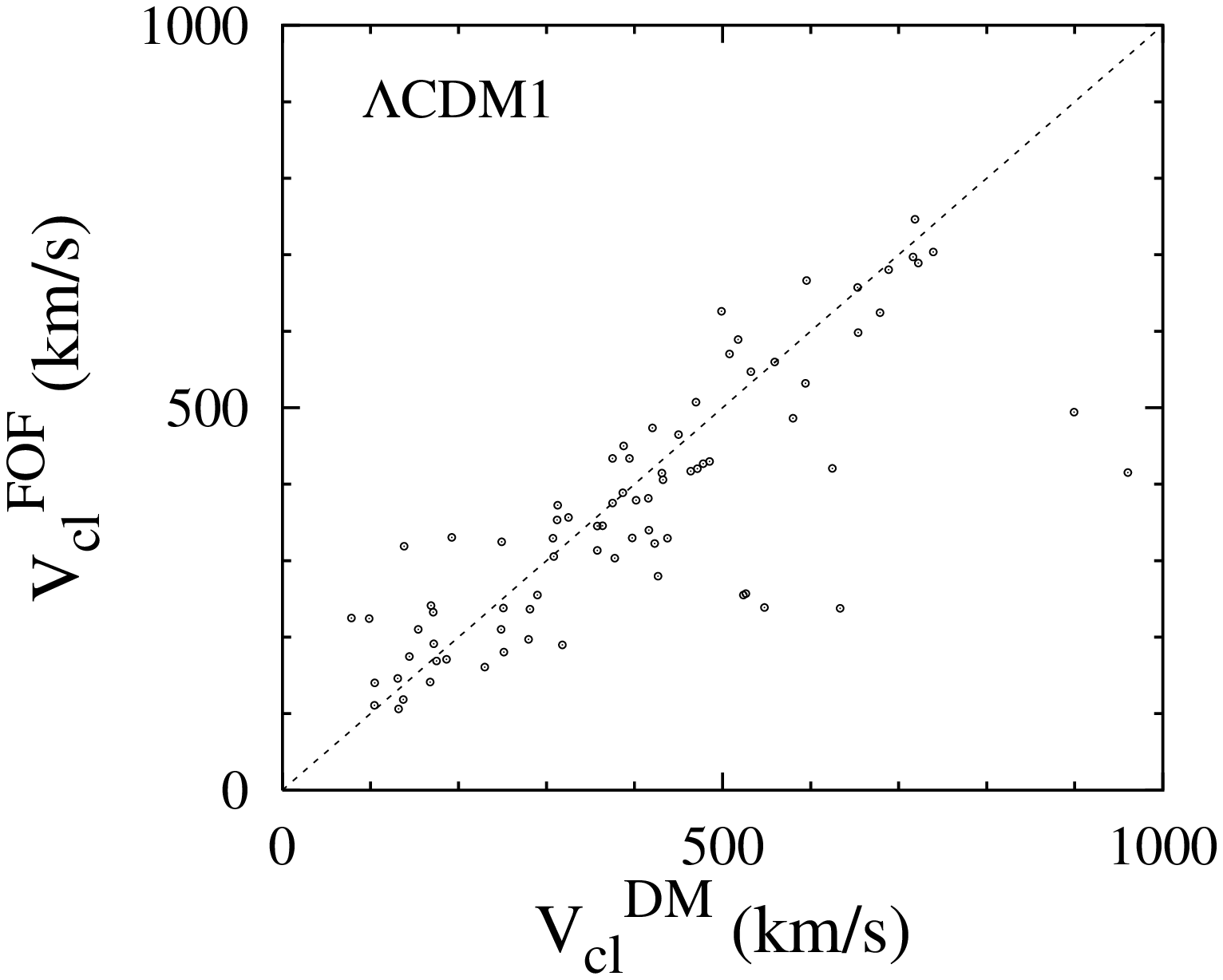,width=8cm}
\psfig{file=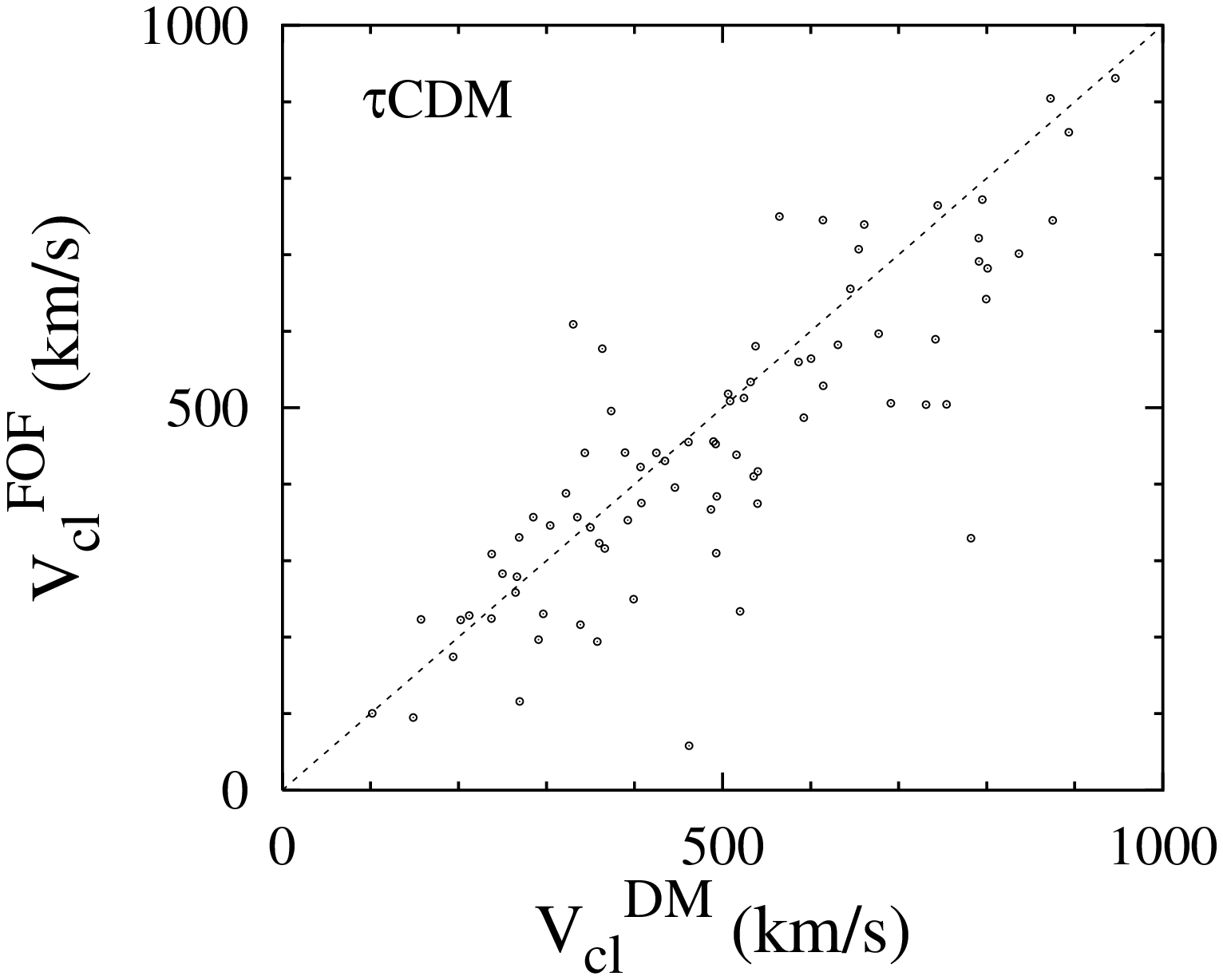,width=8cm}

\caption{ The cluster peculiar velocity defined using the DENSMAX
method vs the cluster peculiar velocity defined by the FOF
method. We show the velocities for rich clusters for
$d_{cl}=50h^{-1}$Mpc. The upper panel shows the velocities in the
$\Lambda$CDM model and the lower panel in the $\tau$CDM model.}
\end{figure}

\sec{SUMMARY AND DISCUSSION}

In this paper we have examined the rms peculiar velocities of
galaxy clusters, $v_{rms}$, for different cluster masses and
radii. We analyzed clusters in the Virgo simulations for two
cosmological models, $\Lambda$CDM and $\tau$CDM  (Jenkins et al.
1998). These simulations were carried out using the AP$^3$M code.
We used the simulations where the mean particle separation
$\lambda_p\sim 1h^{-1}$ Mpc. We also analyzed clusters in an
N-body simulation where the evolution was followed using a PM
code with the same mass resolution that was used in the Virgo
simulations. The cosmological parameters for this simulation were
chosen similar to the Virgo $\Lambda$CDM model. We found that the
rms velocities of clusters in the PM simulation are similar to the
rms velocities in the AP$^3$M simulation. We can use the PM code
to study peculiar velocities on scales that are close to the mean
particle separation $\lambda_p$.

To identify clusters in the simulations we used two methods: the
standard friends-of-friends (FOF) method and the method where the
clusters are defined as maxima of the density field smoothed on
the scale $R \sim 1h^{-1}$ Mpc (DENSMAX). The velocity of DENSMAX
clusters was determined using the same smoothing scale as for the
density field. If we use the DENSMAX method, the size of the
selected clusters is similar for all clusters. The velocity of
FOF clusters was defined to be the mean velocity of all the
particles in the cluster. If we use the FOF method, the size of
the high-mass clusters is larger than the size of the low-mass
clusters. We studied the velocities of FOF clusters defined by
the parameters $b=0.15$, $b=0.2$ and $b=0.3$ (the parameter $b$
defines the linking length as $b\lambda_p$).

We found that the rms velocity of clusters defined with the DENSMAX
method is almost independent of the density of clusters.
The rms velocity of FOF clusters decreases with the cluster mass and radius.
The effect of the cluster radius on $v_{rms}$ is similar to the effect of
the cluster mass on $v_{rms}$ (see Fig.~4 and Fig.~5). The rms velocity of
small clusters is higher than the rms velocity of large massive
clusters. For different values of $b$, this effect is similar.
In the $\Lambda$CDM model, the rms peculiar velocity of massive
clusters with an intercluster separation $d_{cl}=50h^{-1}$ Mpc is
$\approx 15$\% smaller than the rms velocity of clusters with a
separation $d_{cl}=10h^{-1}$ Mpc.

The rms velocity of massive FOF clusters is smaller than the rms
velocity of high-density clusters determined with the DENSMAX
method. What is the reason for this difference? Do we select
different objects by using different methods or do we define
different velocities for the same clusters? We analyzed the
fraction of DENSMAX clusters, $F$, which match FOF clusters (in
terms of their positions). We studied the FOF clusters determined
with $b=0.2$. In the $\Lambda$CDM model, if we compared DENSMAX
clusters for $d_{cl}=50h^{-1}$Mpc with the FOF clusters for
$d_{cl}=50h^{-1}$Mpc, we found that the fraction of matched
clusters is $F=0.72$. For $d_{cl}=30h^{-1}$Mpc in both methods,
the fraction $F=0.74$. In the $\tau$CDM model, the fraction $F$
was similar. We select the same objects by using different
methods, but we rank them in a somewhat different way. In the
DENSMAX method, the clusters are ranked according to their
density and in the FOF method according to their mass. If we
compared DENSMAX clusters for $d_{cl}=50h^{-1}$Mpc with the FOF
clusters for $d_{cl}=40h^{-1}$Mpc, we found that $F=0.98$.

But we do assign different velocities for the same clusters by
using different methods. Fig.~7 shows the peculiar velocities
determined with different methods for the same clusters. We show
the velocities of rich clusters that match for $d_{cl}=50h^{-1}$
Mpc in both methods. We see that the velocities of clusters
defined by the FOF method are systematically smaller than the
velocities of clusters defined by the DENSMAX method. This is
probably caused by the fact that in the FOF method, the size of
rich clusters is larger than the smoothing scale $R\sim 1h^{-1}$
Mpc that was used to define the cluster velocities in the DENSMAX
method.

We compared the rms velocities of clusters with the linear theory
predictions for the rms peculiar velocities of peaks, $\sigma_p$,
for the smoothing radius $R=1h^{-1}$ Mpc. At this radius,
$\sigma_p=509$ kms$^{-1}$ and $\sigma_p=562$ km s$^{-1}$ in the
$\Lambda$CDM and $\tau$CDM models, respectively. We analyzed the
rms velocity of FOF clusters for different values of the cluster
separation. In the $\Lambda$CDM model, the rms peculiar velocity
of small clusters is close to the linear theory expectations,
while the rms peculiar velocity of rich clusters is smaller
($\approx$15 per cent for clusters with a mean separation
$d_{cl}=50h^{-1}$Mpc). The rms velocity of DENSMAX clusters is
almost independent of the cluster density and is similar to the
linear theory expectations. For the DENSMAX clusters with a mean
separation $d_{cl}=30h^{-1}$ Mpc, $v_{rms}=510 \kms$ and
$v_{rms}=580 \kms$ in the $\Lambda$CDM and $\tau$CDM models,
respectively. On scales probed by galaxy clusters, velocity
fluctuations are in the quasi-linear regime.

\sec*{ACKNOWLEDGEMENTS}

We thank J. Einasto, M. Einasto, P. Heinam\"aki, G. H\"utsi, A.
Melott and E. Saar for useful discussions. This work has been
supported by the ESF grant 3601. The N-body simulations used in
this paper are available at {\it
http://www.mpa-garching.mpg.de/Virgo/virgoproject.html.} These
simulations were carried out at the Computer Center of the
Max-Planck Society in Garching and at the EPCC in Edinburgh, as
part of the Virgo Consortium project. The FOF programs used in
this paper are available at {\it
http://www-hpcc.astro.washington.edu.} These programs were
developed in the University of Washington.

\vfill
\end{document}